\documentclass[letterpaper]{sig-alternate}

\usepackage{mathptmx}
\usepackage{ifthen}
\usepackage{epsfig}
\usepackage{color}
\usepackage{amssymb}
\usepackage{setspace}
\usepackage{subfigure}

\newtheorem{definition}{Definition}

\begin{document}
\conferenceinfo{MSWiM'09,}{October 26--29, 2009, Tenerife, Canary Islands, Spain}
\CopyrightYear{2009}
\crdata{978-1-60558-616-9/09/10}

\title{Generalized Analysis of a Distributed Energy Efficient Algorithm for Change Detection}

\numberofauthors{2} 
\author{
\alignauthor
Taposh Banerjee\\
       \affaddr{Dept. of ECE, Indian Institute of Science}\\
       \affaddr{Bangalore, India}\\
       \email{taposh@gmail.com}
\alignauthor
Vinod Sharma\\
       \affaddr{Dept. of ECE, Indian Institute of Science}\\
       \affaddr{Bangalore, India}\\
       \email{vinod@ece.iisc.ernet.in}
}

\maketitle
\begin{abstract}
An energy efficient distributed Change Detection 
scheme based on Page's CUSUM algorithm was presented in \cite{icassp}. In this paper we consider a nonparametric version of this algorithm. In the algorithm in \cite{icassp}, each sensor runs CUSUM and transmits 
only when the CUSUM is above some threshold. The transmissions from the sensors
are fused at the physical layer. The channel is modeled as a 
Multiple Access Channel (MAC) corrupted with noise. 
The fusion center performs another CUSUM to detect the change. In this paper, we generalize the algorithm to also include nonparametric CUSUM and provide a 
unified analysis.
\end{abstract}
\vspace{-0.3cm}
\vspace{-0.3cm}
\vspace{-0.2cm}

\section{Introduction}


In the problem of distributed change detection, there are multiple geographically distributed sensors, 
each sensing a sequence of observations. The distribution of the observations of all the sensors changes 
simultaneously at some random point of time, and the observations are independent and 
identically distributed (\emph{iid}) conditioned on the time of change.
The sensors send processed version (e.g., scaled or quantized) of their observations to a 
decision maker (fusion center) and the fusion center fuses the information from various sensors to detect 
the change of law as soon as possible. The performance metric is a 
measure of the number of extra observations taken after the distribution has changed under some false alarm
constraint (the constraint corresponds to the cost associated with declaring a change when it has 
not taken place). This model finds application in biomedical signal processing, intrusion detection
in computer and sensor networks (\cite{veeravalli01}, \cite{veeravalliISIF03}), finance, quality control engineering, and recently, distributed detection of the primary in cognitive radio networks (\cite{Lataief}, \cite{akjaya}).

In the \emph{Bayesian} formulation of the change detection problem, where the distribution of the change
variable is known, the objective is to minimize the mean delay of detection subject to probability of false alarm. 
The optimal solution is obtained in \cite{Shiryaev63} when the sensors transmit the raw observations
(and hence a centralized solution is feasible), and in 
\cite{veeravalli01} when the sensors transmit a quantized version of the observations. 

In the \emph{Min-Max} formulation, when no knowledge is assumed of the distribution of change, the worst case delay is minimized subject to a constraint on the mean time to false alarm. When the sensors send the observations in raw, the CUSUM algorithm (first proposed by Page in \cite{Page}) was shown to be optimal in \cite{lorden71} and  \cite{Moustakides86}. When the sensors process the information before transmitting, CUSUM based schemes were shown 
to be optimal in \cite{Mei}.

The above algorithms minimize delay subject to constraints on only false alarm. Hence, it is not clear
if they are also energy efficient. As a result, the distributed algorithm which optimizes delay 
under any of the above two formulations, under \emph{both} false alarm and energy constraints, is not yet 
known. Recently, a Bayesian formulation of the decentralized change detection 
problem with energy constraints was considered in \cite{LeenaRajesh}. The problem is solved 
by restricting the solution to a class of algorithms where the sensors send scaled and shifted versions of 
their observations to the fusion center. The algorithm in \cite{LeenaRajesh} is energy efficient 
but leaves out a class of algorithms where the sensors can decide not to transmit based on the available 
information. This latter technique can be used to save more energy. 

This issue is exploited in \cite{icassp} to propose a CUSUM based algorithm called DualCUSUM. This new 
algorithm was shown to perform much better than the one in \cite{LeenaRajesh} and the gap in performance 
was shown to increase at lower energy and false alarm constraints.

In this paper we provide the false alarm and delay analysis of DualCUSUM. We also 
extend DualCUSUM to detect changes when the exact information about the 
pre-change and post-change distributions is not available (nonparametric setting) and provide a 
unified analysis. 


As will be seen from the paper, the analysis of such distributed cooperative systems is very challenging. 
We use results from Renewal Theory, Extreme Value Theory and theory of Brownian Motion to
analyse our algorithm. Since CUSUM is, or will be a fundamental element of many distributed algorithms
for detection of change, the tools and techniques used here can be used to analyse those other 
algorithms as well. 

DualCUSUM has been used for cooperative spectrum sensing in Cognitive Radio 
Systems also in \cite{akjaya} and shown to provide better performance than other algorithms 
available in literature not only in delay in detection but also in saving energy. 


The knowledge of pre-change and post-change distributions may not be practical in many cases: 
random time varying fading in the wireless channels (\cite{ChamberLand}, \cite{Lataief}) and due to lack 
of information related to transmit power and/or receiver noise (\cite{sahai}).
Thus in this paper we also extend DualCUSUM to a nonparametric set up. 

We analyse the generalized version of DualCUSUM 
of which parametric and nonparametric versions are special cases. A few interesting facts emerge from this analysis: mean detection delay is independent of the distribution of the observations but the false alarm probability crucially depends on the tail behavior of the distributions (at least for the nonparametric case). 
The lighter the tail, the lower the false alarm probability. Therefore for a given false alarm constraint, under both parametric and nonparametric settings, a system where the observations are lighter tailed, will have lower detection delay. 

We also show that the log likelihood function converts a heavy tailed distribution to a light tail 
distribution, or in general, makes the tail of the distribution lighter. 
Since, parametric CUSUM uses log likelihood and nonparametric CUSUM 
does not, the former performs better than the latter for a given distribution of observations.
This interesting property of log likelihood and its implications for CUSUM seem to have gone unnoticed so far.



The paper is organized as follows. 
We explain the model and introduce the algorithm in Section \ref{sec:DetectionOverview}. 
Section \ref{sec:Analysis} analyzes the performance of the algorithm and provides comparison
with simulations. Section \ref{sec:conclusion} concludes the paper. 
\vspace{-0.7cm}
\vspace{0.1cm}
\section{Model and Algorithm}
\label{sec:DetectionOverview}
Let there be $L$ sensors in a sensor field, sensing observations and transmitting to a fusion 
node. The transmissions from the sensor nodes to the fusion node are via a MAC. In our system we 
assume that all the sensor nodes can transmit at the same time. There is physical layer 
fusion at the fusion node (commonly studied Gaussian MAC is a special case). 
The fusion node receives data over time and decides if there is a change 
in distribution of the observations at the sensors. 

Let $X_{k,l}$ be the observation made at sensor $l$ at time $k$.
Sensor $l$ transmits $Y_{k,l}$ at time $k$ after processing $X_{k,l}$
and its past observations. 
The fusion node receives $Y_k = \sum_{l=1}^L Y_{k,l} + Z_{MAC,k}$, where $\{Z_{MAC,k}\}$ is 
iid receiver noise. 
The distribution of the observations at each sensor changes simultaneously 
at a random time $T$, with a known distribution. 
The $\{X_{k, l}, l\geq 1\}$ are independent and identically distributed (iid) over $l$ and are 
independent over $k$, conditioned on change time $T$. 
Before the change $X_{k,l}$ have density
$f_0$ and after the change the density is $f_1$. 
The expectation under $f_i$ will be denoted by $E_i$, $i=0, 1$,
and $P_\infty$ and $P_1$ denote the probability measure 
  under no change and when change happens at time 1, respectively.
These assumptions are commonly made in the literature 
(see e.g., \cite{Mei} and \cite{veeravalli01}) 

The objective of the fusion center 
 is to detect this change as soon as possible at time $\tau$ (say)
 using the messages transmitted from all the $L$ sensors, subject to 
an upper bound $\alpha$ on the probability of False Alarm $P_{FA} \stackrel{\triangle}{=} P(\tau < T)$
and ${\cal E}_0$ on the average energy used. Often the desired $\alpha$ is quite low, e.g., $ \leq 10^{-6}$ in 
intrusion detection in sensor networks.
Then, the general problem is:
\begin{eqnarray}
\label{eq:detectionProblem}
\min   E_{DD} \hspace{-0.3cm}&\stackrel{\triangle}{=}&\hspace{-0.3cm} E[(\tau-T)^+], \nonumber \\
\mbox{ Subj to }  P_{FA} \hspace{-0.3cm}&\leq& \hspace{-0.3cm}\alpha \mbox{ and } {\cal E}_{avg}  \hspace{-0.1cm} =   E\left[\sum_{k=1}^\tau Y_{k,l}^2\right] \leq {\cal E}_0,  1\leq l \leq L.
\end{eqnarray}
  Our algorithm in \cite{icassp} does not provide an optimal solution to 
the problem but uses several desirable features to
 provide better performance than the algorithms we are aware of. 
We reproduce the algorithm, DualCUSUM, for an easy reference:

\begin{enumerate}
\item Sensor $l$ uses CUSUM,  
\begin{eqnarray}
\label{SensorCUSUM}
W_{k,l} =  \max \left(0, W_{k-1, l} + \log \left[f_1\left(X_{k,l}\right)\left / f_0\left(X_{k,l} \right. \right) \right] \right),  
\end{eqnarray}
where, $W_{0,l}  =   0, 1\leq l \leq L$.
\vspace{-0.2cm}
\item Sensor $l$ transmits $Y_{k,l} = b 1_{\{W_{k,l}> \gamma\}}.$
Here $1_A$ denotes the indicator function of set A and $b$ is a design parameter. 
\vspace{-0.2cm}
\item Physical layer fusion (as in \cite{LeenaRajesh}) is used to reduce transmission delay, i.e., 
$Y_k = \sum_{l=1}^L Y_{k,l} + Z_{MAC, k}$, where $Z_{MAC, k}$ is the receiver noise. 
\vspace{-0.2cm}
\item Finally, Fusion center runs CUSUM:
\begin{eqnarray}
\label{eq:FuseCUSUM}
 F_k = \max \left\{0, F_{k-1} + \log \frac{g_I(Y_k)}{g_0(Y_k)} \right\}; \  \ F_0=0,
\end{eqnarray}
where $g_0$ is the density of $Z_{MAC,k}$, the MAC noise at the fusion node, and $g_I$ is the density
of $Z_{MAC,k}+bI$, $I$ being a design parameter. 
\item The fusion center declares a change at time $\tau(\beta, \gamma, b, I)$ 
when $F_k$ crosses a threshold $\beta$: $\tau(\beta, \gamma, b, I) =  \inf \{ k: F_k > \beta \}.$
\end{enumerate}

Multiple values of ($\beta, \gamma, b, I$) will satisfy both the false alarm 
and the energy constraint. One can minimize $E_{DD} \stackrel{\triangle}{=} E[(\tau-T)^+]$ over this parameter set. 
 In Section \ref{sec:Analysis}
we obtain the performance of DualCUSUM for given values $(\beta, \gamma, b, I)$
which then can be used in the optimization algorithm developed in \cite{icassp} to 
efficiently solve the optimization problem above. 

 DualCUSUM, as the original CUSUM itself, has a strong limitation. It requires exact knowledge
of $f_0$ and $f_1$. This information will be available apriori to varying degrees 
in a practical scenario. 
Depending upon the type of uncertainty in $f_0, f_1$, different algorithms/variations
on CUSUM are available (\cite{Brodsky}, \cite{lai95}). 
One common algorithm, called nonparametric CUSUM is to replace 
(\ref{SensorCUSUM}) by
\begin{eqnarray}
\label{eq:NonParametricVer}
W_{k+1,l} = \max \{0, W_{k,l} + X_{k,l} - D\},
\end{eqnarray}
where, $D$ is an appropriate constant such that $E[X_{k,l} - D]$ is negative
before change and positive after change. 
If the mean of $X_{k,l}$ is known before and after the change, $D$ can be chosen 
as the average of the two means. 
Depending on the uncertainties involved, we should be able to obtain such a $D$. 
More generally, we should be able to estimate $D$. 
For Gaussian and exponential distributions, nonparametric CUSUM becomes CUSUM 
for some appropriate $D$ and scaling.
If at the fusion node $g_0$ is not known (in our CUSUM algorithm (\ref{eq:FuseCUSUM}) at the 
fusion node, $g_I(x) = Ib + g_0(x)$), then one can use (\ref{eq:NonParametricVer})
even at the fusion node. 

In the following we will compute $P_{FA}$ and $E_{DD}$ for a generalized class of algorithms 
where at the sensor nodes and at the fusion node we use the algorithm,
\begin{eqnarray}
\label{eq:GeneralCUSUM}
W_{k+1} = \max \{0, W_k + Z_{k+1}\},
\end{eqnarray}
where, $\{Z_k\}$ is an iid sequence with different distributions before and after the change. 
We will assume that $E[Z_k]<0$ before the change and $E[Z_k]>0$ after the change. 
We will denote by $f_Z, F_Z$ and $P_Z$ the density, cdf and probability measure for $Z_k$. 

Algorithm (\ref{eq:GeneralCUSUM}) contains CUSUM and nonparametric CUSUM as special cases. In the
next section we analyze the generalized DualCUSUM with (\ref{eq:GeneralCUSUM}). We emphasize that
unlike DualCUSUM, this algorithm \emph{does not} require knowledge of $f_0$ and $f_1$; we only need 
to chose $D$ appropriately. But, the performance of this algorithm, as we show in the next section, does 
depend on the underlying distributions. This is typical of such algorithms. 

\section{Analysis}
\label{sec:Analysis}
\begin{figure}[htbp]
\centering
\input{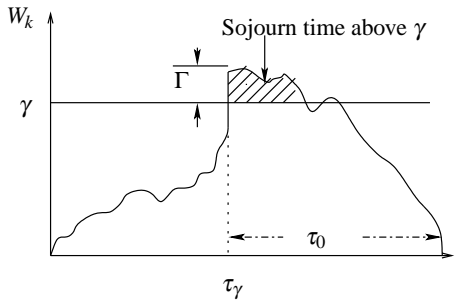}
\caption{{\scriptsize Excursions of $W_k$ above $\gamma$ can be 
         approximated by a compound Poisson process. A local node transmits to the fusion node 
         during these excursions.}}
\label{fig:CUSUMEvolution}
\end{figure}

In this section, we first compute the false alarm probability $P_{FA}$ and then the delay $E_{DD}$. 
The idea is to model the times at which the CUSUM $\{W_k\}$ at the local sensors,  
crosses the threshold $\gamma$ (we drop subscript $l$ for convenience)
and the local nodes transmit
 to the fusion node (Fig. \ref{fig:CUSUMEvolution}). 

Computing $P_{FA}$ requires finding (when $Z_k$ has distribution $f_0$) 
the distribution of $\tau_\gamma$, the first time $W_k$ crosses $\gamma$, the amount of time 
it stays above $\gamma$ (sojourn time above $\gamma$), and the probability that the fusion node 
declares a change during a sojourn time. These are computed in Sections 
\ref{sec:Analysis_sensor}-\ref{sec:Analysis_falseAlarm}.
Computation of $E_{DD}$ is sketched in Section \ref{sec:Analysis_Edd}.

We will need the following notations and definitions. Let $X$ be a random variable with distribution $F$. 
Then $F^{*n}$ denotes the $n$-fold
convolution of $F$ and $\bar{F}(x) = 1-F(x)$.
\begin{definition}
\label{def:HeavyTail}
(\cite{asmussen98}) $F$ is \textbf{heavy tailed} if for any $\epsilon > 0$, $E[ e^{\epsilon |X|} ] = \infty$.
$F$ is \textbf{subexponential} if 
$\bar{F}^{*2}(x)/\bar{F}(x) \to 2  \mbox{ as }  x\to \infty.$
If $F$ is not heavy tailed, we call it \textbf{light tailed}. 
\end{definition}

 Gaussian, Exponential and Laplace distributions are light tailed. 
Pareto, Lognormal and Weibull distributions are subexponential.
Subexponential distributions are a subclass of heavy tailed distributions.

Often it is said that light tailed distributions may provide better system behavior 
than the heavy tailed (\cite{Boxma}). We demonstrate this for the probability of false alarm.
In particular we will show that if $F_Z$ is light tailed before change then $P_{FA}$ is much less than if it
is heavy tailed. Interestingly, we will also show that $E_{DD}$ is largely insensitive to the tail 
behavior of $F_Z$.

CUSUM has the interesting property that it transforms a large class of heavy tailed distributions into  light tailed distributions. 
Consider the class of Pareto distributions:
\[K \frac{x_m^K}{x^{(K+1)}} \mbox{ for } x\geq x_m, \mbox{ where } x_m>0.\]
For $f_0$, we take $K=(\alpha+1)$ and for $f_1$, $K=\alpha$ where $\alpha $
is a positive constant.
Then, if $Z=log(f_1(X)/f_0(X))$, $ P_\infty[ Z_k > z ]  \sim e^{-z\alpha}$. On the other hand, 
light tailed distributions stay light tailed. 
This important property 
of log likelihood seems to have escaped the attention of investigators before.
This makes CUSUM perform better than the nonparametric CUSUM because as mentioned above, 
for CUSUM with Pareto distribution, the $P_{FA}$ will be as for a light tailed distribution. 
But for nonparametric CUSUM, it will be much larger, while the $E_{DD}$ in the two cases will be 
comparable.

\subsection{Behavior of {\large $\textbf{\emph{W}}_{\textbf{\emph{k}}}$ \textbf{under} 
$\textbf{\emph{P}}_{\mathbf{\infty}}$}}
\label{sec:Analysis_sensor}
The process $\{W_k\}$ is a reflected random walk with negative drift under $P_\infty$.
Figure (\ref{fig:CUSUMEvolution}) shows a typical sample path for $\{W_k\}$. 
The process visits 0 (regenerates) a finite number of times 
before it crosses the threshold $\gamma$ at, 
\begin{eqnarray}
\label{eq:FPT}
 \tau_\gamma  \stackrel{\triangle}{=} \inf \{ k \ge 1 : W_k \ge \gamma \}.
\end{eqnarray}
We call $\tau_\gamma$  the \textit{First Passage Time (FPT)}. 
The \emph{overshoot} $\Gamma = W_{\tau_\gamma} - \gamma$. 
Let \begin{eqnarray}
\label{eq:etaDefine}
 \tau_0 &\stackrel{\triangle}{=}& \inf \{k: k>\tau_\gamma ; W_k \leq 0 \}-\tau_\gamma \hspace{1.00cm} \mbox{  and  } \nonumber \\  
 \eta &=& \# \{k: W_k \geq \gamma; \tau_\gamma \leq k \leq \tau_\gamma + \tau_0\}.
\end{eqnarray}

During time $\eta$ (called a batch) a local node transmits to the fusion node. Thus, 
these are the times during which the fusion node will most likely declare a change. 
The overshoot $\Gamma$ can have significant impact on $\eta$.

It has been shown in \cite{Rootzen} that the point process of exceedances of $\gamma$
by $W_k$, converges to a compound Poisson process as $\gamma \to \infty$. 
The points appear as clusters. The intervals between the clusters have the same distribution as that of  
$\tau_\gamma$ in (\ref{eq:FPT}) and the distribution of $\eta$ in (\ref{eq:etaDefine})
gives the distribution of the size of the cluster, i.e., the batch of the compound Poisson process.  
Since, one has to choose large values of $\gamma$ to keep $P_{FA}$ small, a batch Poisson process
 provides a good approximation in our scenario.

In the next few sections we give results on the distribution of $\tau_\gamma$, 
overshoot $\Gamma$, and the distribution of the batch $\eta$ which 
will be used in computing $P_{FA}$. 

\subsection{First Passage Time under {\large $\textbf{\emph{P}}_{\mathbf{\infty}}$}} 
\label{sec:FPT}
From the compound Poisson process approximation mentioned above,
\begin{eqnarray}
\label{eq:expoSensor} 
\lim_{\gamma \to \infty} \mathbf{P}_{\infty} \{\tau_\gamma >  x\} = \exp(-\lambda_\gamma x), \mbox{   } x>0,
\end{eqnarray}
where, $\lambda_\gamma$ a positive constant. 

In \cite{icassp} a formula for $\lambda_\gamma$ was used which is computable for Gaussian distribution only. 
However, by solving integral equations obtained via renewal arguments (\cite{Ross}), one can obtain  
the mean of FPT for any distribution. Epochs when $W_k=0$ are renewal epochs for this process. 
Let $L(s)$ be the mean FPT with $W_0=s\geq 0$. Hence $\lambda_\gamma=1/L(0)$. Then from renewal arguments:
\begin{eqnarray}
\label{eq:FPTIntegralEqn}
L(s) = F_Z(-s)L(0) + \int_{-s}^\gamma L(s+z) dF_Z(z) dz + P[Z>\gamma-s].
\end{eqnarray}
The equation is obtained by conditioning on $Z_0=z$. If $Z_0 \leq -s$, then $W_1=0$, providing the 
first term on the right. If $Z > \gamma - s$, then the threshold is approached in one step only, 
providing the last term. This equation can be solved recursively on $L(s), 0\leq s \leq \gamma$.
An algorithm provided in \cite{LucenoARLDist} can be used to compute 
 (\ref{eq:FPTIntegralEqn}) efficiently. Table \ref{Tab:FPT} provides $E[\tau_\gamma]$ for 
Pareto distribution with $K=2.1$ and Gaussian distribution with $EZ_k = -0.5$ and $var(Z_k)=1$.
We see that as $\gamma$ increases $E[\tau_\gamma]$ for Gaussian distribution becomes much larger than for 
the Pareto distribution. This implies that $P_{FA}$ for the Gaussian distribution should be much less than 
for Pareto, $K=2.1$ if $\gamma$ is large. 
\begin{figure}[htbp]
\centering
{\scriptsize 
\begin{tabular}{|r|r|r|r|r|r|r|r|r|r|r|}
\hline 
$\gamma$&$E[\tau_\gamma]$ \mbox{Gauss}&$E[\tau_\gamma]$ \mbox{Pareto}	\\
\hline
5&930&800\\
6&2551&1100\\
7&6950&1455\\
8&19020&1880\\
\hline
\end{tabular}} 
\makeatletter\def\@captype{table}\makeatother
\caption{{\scriptsize Mean FPT $E[\tau_\gamma]$ for Pareto ($K=2.1$) and Gaussian with $EZ_k=-0.5$. }}
\label{Tab:FPT}  
\end{figure}
\subsection{Distribution of overshoot}
Next we consider mean and distribution of the overshoot $\Gamma$. From renewal equations as in 
(\ref{eq:FPTIntegralEqn}), we can exactly compute $E\Gamma$ for any distribution.
If $R(x) = E\Gamma$ with $W_0 = x$, 
\begin{eqnarray}
\label{eq:OverShootIntEqn}
R(x) & = & E[Z_k-(\gamma-x) | Z_k>(\gamma-x)] P_Z(Z_k>\gamma-x) \nonumber \\
&+& \int_{y=0}^\gamma R(y) f_Z(y-x) dy + R(0) F_Z(-x).
\end{eqnarray}
Mean overshoot $E\Gamma = R(0)$.
For light tails $E\Gamma$ converges quickly to a constant value as $\gamma \to \infty$. 

The distribution of the overshoot for heavy tails is reported in \cite{asmussen98}.
One can show that for light tailed distributions, overshoot is exponentially distributed. 
One can also obtain exact distribution using renewal integral equations as in (\ref{eq:FPTIntegralEqn})
and (\ref{eq:OverShootIntEqn}).
\begin{figure}[h]
\begin{center}
\epsfig{figure=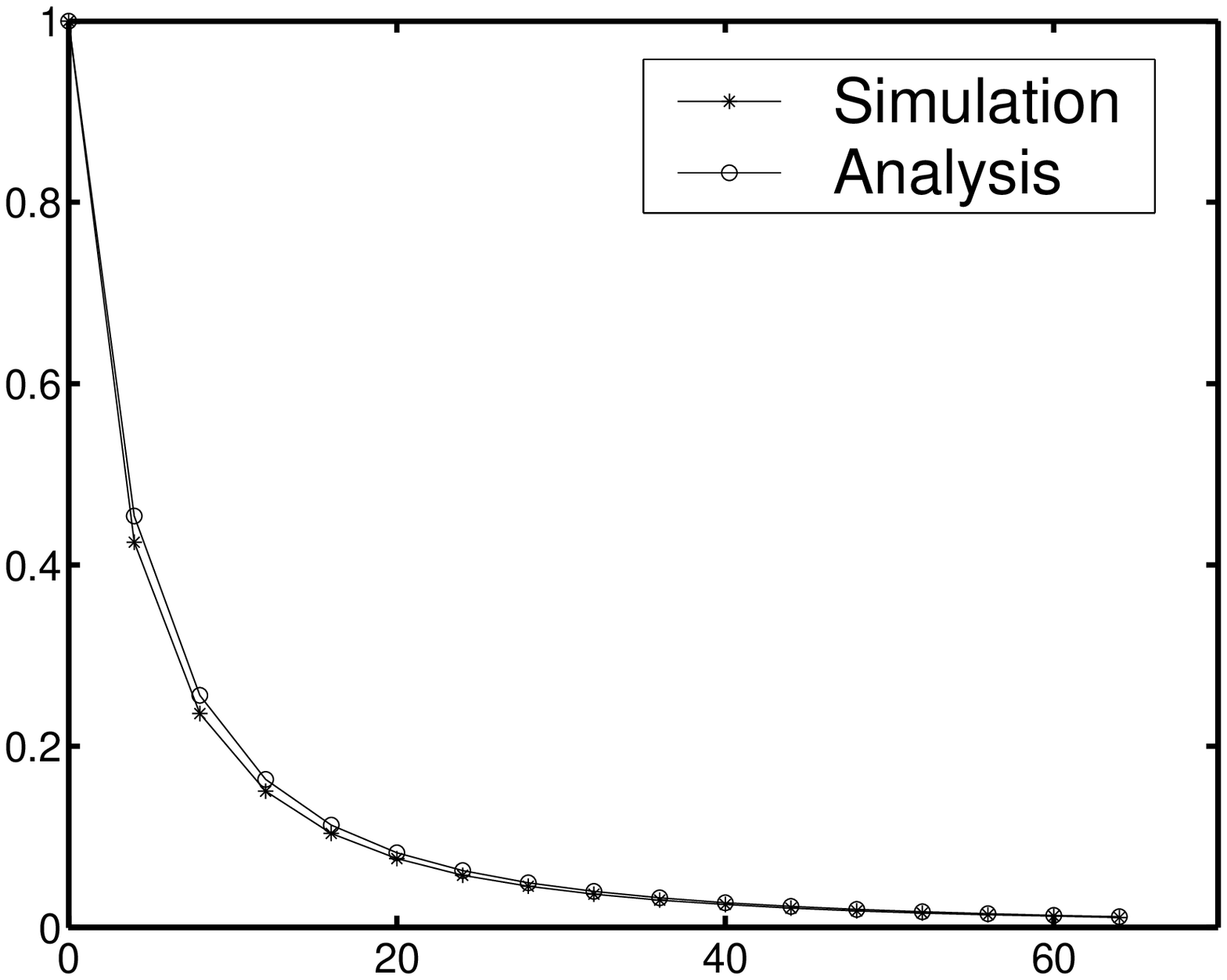,height=4cm,width=5cm}
\caption{{\scriptsize Complementary CDF of $\Gamma$ for Pareto $K=2.1$, $EZ_k=-0.3$ and $var(Z_k)=1$ and $\gamma=8$}}
\label{fig:OvershootDistParetoK2pt1}  
\end{center}
\end{figure}
\begin{figure}[h]
\begin{center}
\epsfig{figure=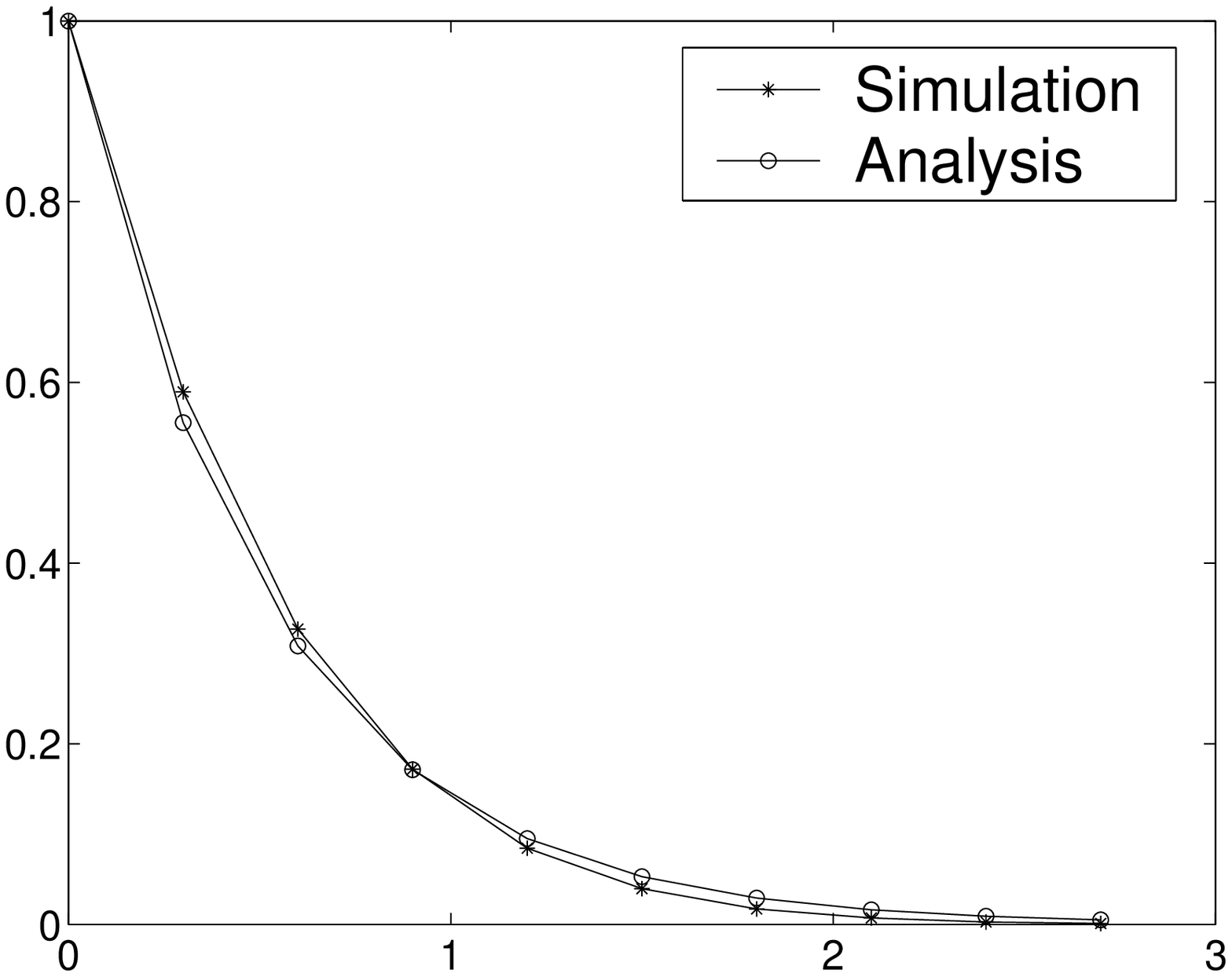,height=4cm,width=5cm}
\caption{{\scriptsize Complementary CDF of $\Gamma$ for $Z_k \sim N(-0.3,1)$ and $\gamma\geq 6$}}
\label{fig:OvershootDistGauss}
\end{center}
\end{figure}
We plot the distribution of overshoot for Pareto distribution with $K=2.1$ in Figure (\ref{fig:OvershootDistParetoK2pt1}) and for Gaussian distribution in Figure (\ref{fig:OvershootDistGauss}). 
The mean overshoot $E[\Gamma]$ 
was obtained using equation (\ref{eq:OverShootIntEqn}). 
The distribution of $\Gamma$ for Pareto distribution was obtained via the result in \cite{asmussen98}
and for the Gaussian distribution by exponential distributions. 
 
 Comparing Figures (\ref{fig:OvershootDistParetoK2pt1}) 
and (\ref{fig:OvershootDistGauss}), we see that the analytical approximations are very good. 
Also, the overshoot for Pareto distribution is much 
more than for the Gaussian distribution.
\subsection{Distribution of the Batch}
\label{sec:BatchDist}
\vspace{-0.3cm}
\subsubsection{Distribution of batch for heavy tail}
Theorem 2.4 of \cite{asmussen98} provides the batch size distribution for subexponential $Z$. 
Figure (\ref{fig:BatchHeavyTailParetoK2pt1})
shows the plot of Batch complementary CDF for Pareto distribution with parameters $K=2.1$. 
One sees a good match with simulations.

\subsubsection{Distribution of batch for light tail}
\begin{figure}[h]
\begin{center}
\epsfig{figure=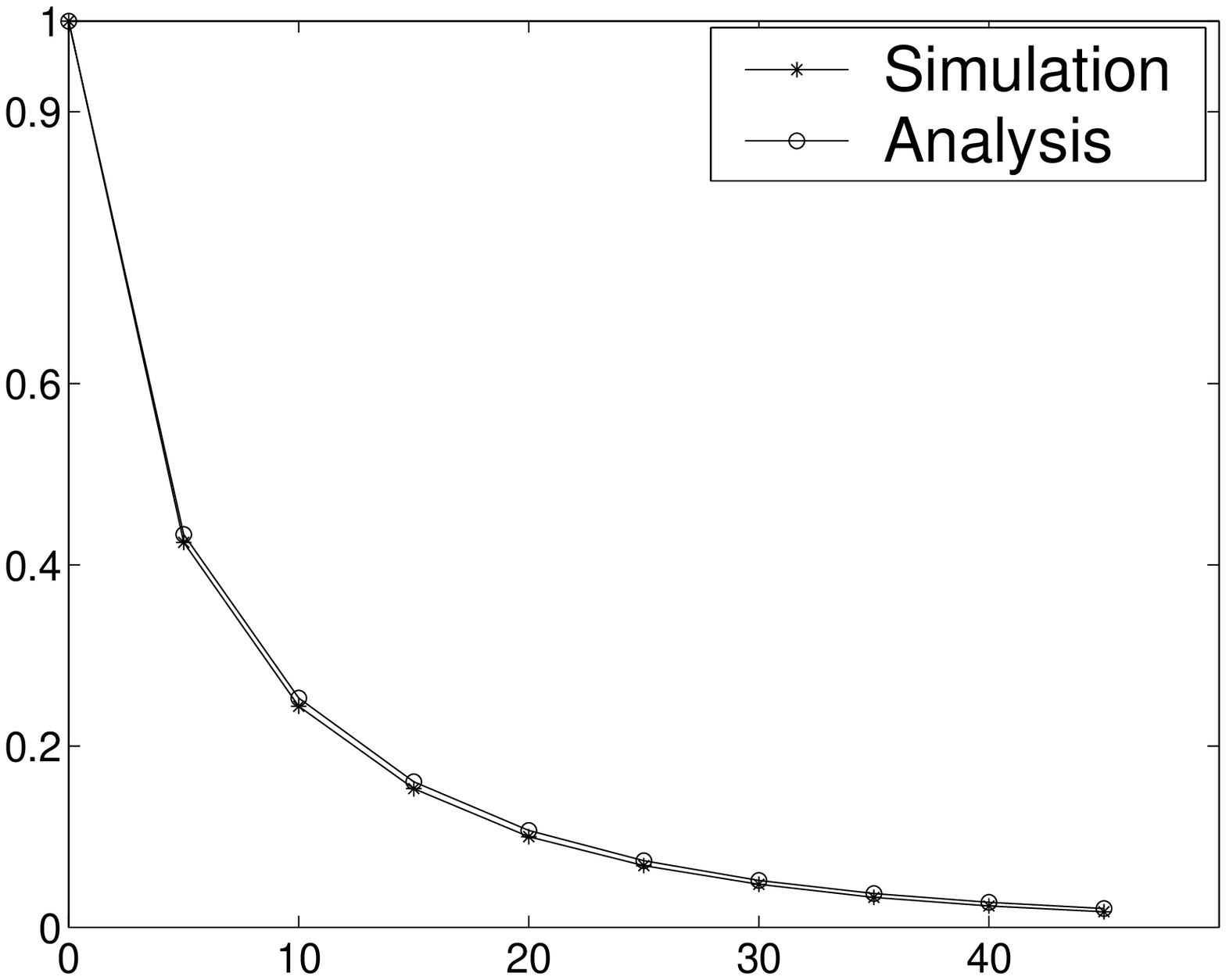,height=4cm,width=5cm}
\caption{{\scriptsize Complementary CDF of Batch $\eta$ for Laplace $Z_k$ with $EZ_k=-0.3$ and $var(Z_k)=1$ 
and $\gamma\geq 7$.  }}
\label{fig:BatchLightLaplace}
\end{center}
\end{figure}

Let $G_j(x)$ be the conditional batch distribution,
$ G_j(x) = P(\eta \leq j | W_{\tau_\gamma} = \gamma + x) ,$
when the overshoot is $x$.
$G_j(x)$ was obtained in \cite{icassp} by the Brownian Motion (BM) approximation of 
$\{W_k\}$. 
We use this approximation along with exponential distribution for $\Gamma$ to obtain 
\begin{eqnarray}
\label{eq:NewBatchDist}
P(\eta \leq j) = \int_{0}^\infty G_j(x) \frac{1}{E[\Gamma]} exp(-\frac{x}{E[\Gamma]}) dx
\end{eqnarray}
for light tailed distributions. 
Figure (\ref{fig:BatchLightLaplace}) plots the distribution of $\eta$ for $Z_k$ with Laplace 
distribution via (\ref{eq:NewBatchDist}) and via simulations.

For Lognormal distribution, which can be approximated via both heavy tailed and light tailed approximations provided above, (\ref{eq:NewBatchDist}) provides a better approximation.

\begin{figure}[h]
\begin{center}
\epsfig{figure=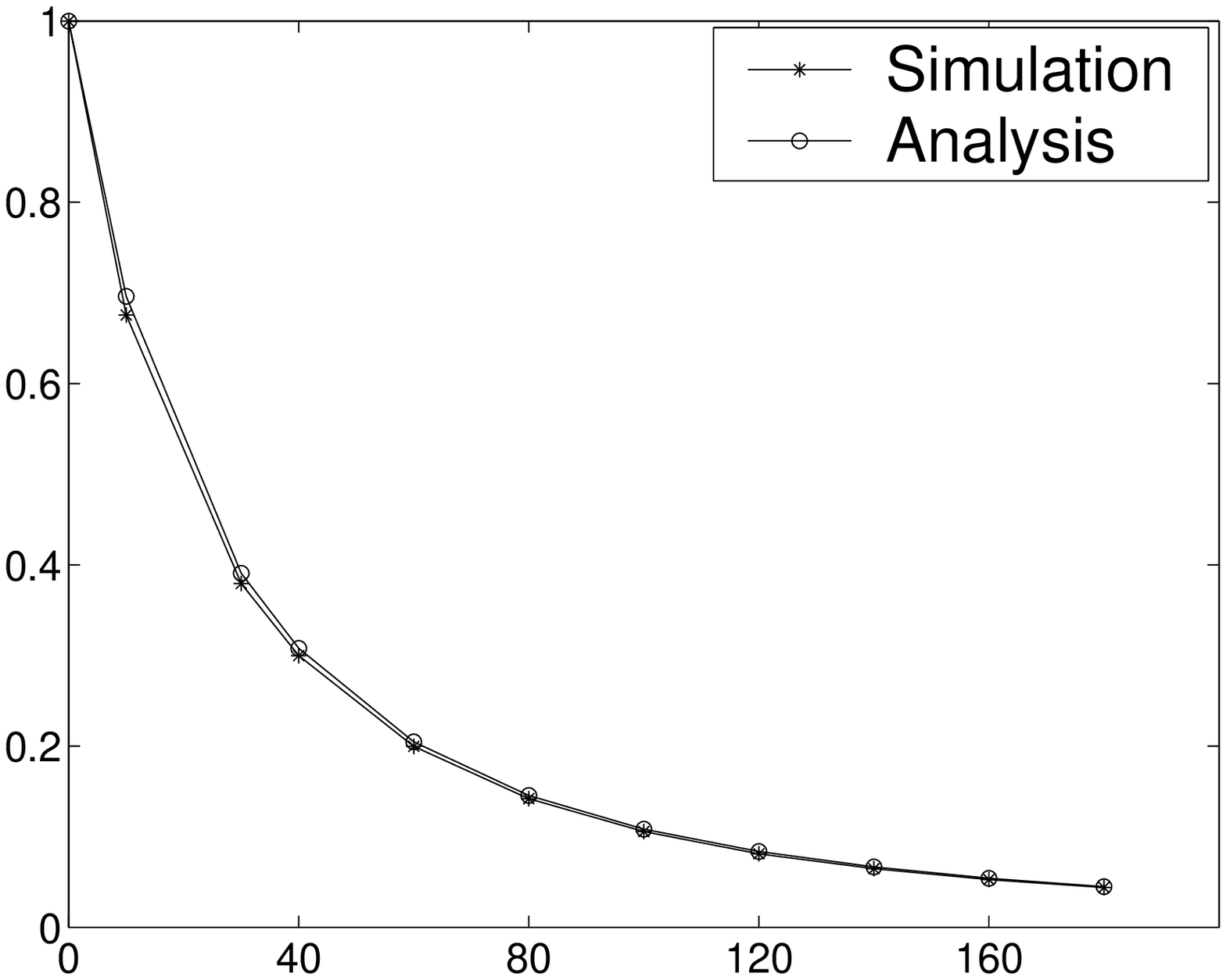,height=4cm,width=5cm}
\caption{{\scriptsize Complementary CDF of Batch $\eta$ for Pareto $K=2.1$, $EZ_k=-0.3$ and $var(Z_k)=1$ 
and $\gamma=15$.}}
\label{fig:BatchHeavyTailParetoK2pt1}
\end{center}
\end{figure}

Comparing Figures (\ref{fig:BatchHeavyTailParetoK2pt1}) and (\ref{fig:BatchLightLaplace})
one sees that the batch size for a Pareto distribution is larger than for a 
Laplace distribution even when they have same mean and variance.  This is a direct 
consequence of having larger overshoots. 

\vspace{-0.2cm}
\subsection{False Alarm Analysis}
\label{sec:Analysis_falseAlarm}
The false alarm in DualCUSUM can happen in two ways: one within a batch 
(we denote its probability by $\tilde{p}$) and another outside it, i.e., due to $\left\{Z_{MAC,k}\right\}$. 
We will compute these later on. Now, we compute the $P_{FA}$ from these quantities. 

From the assumptions made and the above approximation, the inter-arrival time of
  the batches in the system (at the fusion center) is exponentially 
 distributed with rate $L \lambda_{\gamma}$ (because the processes $\{W_{k,l}\}$
 are independent for different nodes each generating batches as Poisson processes with rate $\lambda$). 
Then, the number of batches appearing before the 
time of change is a Poisson random variable with parameter $L \lambda_\gamma i$,
when $T=i$.
In the following, 
we will show that the time to FA outside a batch is exponentially distributed 
with parameter $\lambda_0$ (to be defined below). 
Therefore, if $T \sim \mbox{Geom}(\rho)$, then one can show that:
\begin{eqnarray}
\label{eq:FalseAlarmExpression}
P_{FA} = 1-\frac{e^{-(\lambda_0+\lambda_\gamma L \tilde{p} )} \rho}{1-e^{-(\lambda_0+\lambda_\gamma L \tilde{p})}(1-\rho)}.
\end{eqnarray}
Similarly, one can obtain expression for $P_{FA}$ when $T$ is not geometric. 

\subsection*{False Alarm within a Batch:}

\label{sec:Pfa_IntsideBatch}
We have seen above that for light tailed $Z_k$, the $E[\tau_\gamma]$
is large and the batch sizes are small. Thus, the batches by different local nodes
do not overlap. However, it is not true for heavy tailed distributions. Thus we compute 
$\tilde{p}$ for the two cases separately. \\

\textbf{Light Tailed} 

The false alarm probability ($\tilde{p}$) within a batch, can be computed as, 
$\tilde{p} \approx \sum_{i=1}^\infty P(\eta=i) P(\mbox{FA }| \eta=i )$, where
$P(\mbox{FA }| \eta=i )$ represents the probability of
FA (CUSUM at the fusion center crossing $\beta$) in $i$ transmissions when one local node 
is already transmitting, i.e., $Y_k = b + Z_{MAC, k}$. 
If $\tau_\beta$ is the FPT variable at the fusion center, then, $P(\mbox{FA }| \eta=i ) = P(\tau_\beta \leq i).$
Since $\eta$ is small for negative drift under $f_0$ ($D$ in (\ref{eq:NonParametricVer})
is chosen that way) we use
 integral equations  to compute the distribution of $\tau_\beta$ for 
observations $Y_k$ given in this paragraph. 

Table \ref{Tab:Pfa} gives the comparison of the $P_{FA}$ values 
obtained via (\ref{eq:FalseAlarmExpression}) and simulations for light tailed 
distributions Gaussian and Laplace. It turns out that the expression is also 
valid for heavy tails like Lognormal (also shown in Table \ref{Tab:Pfa}). One can 
see a good match.  \\

\textbf{Heavy Tailed}

 Now, we use different arguments to compute $\tilde{p}$ and then use it in (\ref{eq:FalseAlarmExpression}).
For simplicity, in the following, the fusion center 
 is assumed to use (\ref{eq:FuseCUSUM}) for detection and not nonparametric CUSUM. From \cite{icassp}, 
 the optimal choice of $I$ is found to be always greater than 1. 
 
 Let $m$ be the minimum number of sensors required to make drift of $F_k$
 positive. We denote the drift, with $m$ nodes transmitting, by $\mu_m$. Then 
 we approximate $\tilde{p}$ by the probability that $F_k$ will have positive drift during
 a batch and that the batch lasts for $\beta/\mu_m$ time (the time needed for $F_k$ to cross 
$\beta$ when the drift is $\mu_m$) after $m$ sensors start transmitting. 
We use this approximation to compute $P_{FA}$ for Pareto $K=2.1$ distribution. 
This is also provided in Table \ref{Tab:Pfa}. We see that the approximation is indeed good for Pareto $K=2.1$. 
\begin{table} [htbp]
\centering 
\centering
{\scriptsize
\begin{tabular}{|r|r|r|r|r|r|r|r|r|r|r|}
\hline 
&$L$&$I$&$\gamma$&$\beta$&$ P_{FA} $&$ P_{FA} $\\
&&&&&$Anal.$&$Sim.$\\
&&&&&$\times 10^{-4}$&$\times 10^{-4}$\\
\hline
Gauss&5&2&15&18&1.22&1.1\\
&10&2&15&18&2.43&2.28\\
\hline 
Laplace&6&2&16&16&2.57&2.06\\
&12&3&16&16&0.66&0.55\\
\hline
Log-&5&2&25&20&1.47&1.76\\
normal&10&2&25&20&2.97&3.5\\
\hline 
Pareto&5&3&30&30&1.93&1.77\\
 K=2.1&5&3&50&50&0.23&0.25\\
\hline
\end{tabular} }
\caption{{\scriptsize $P_{FA}$ for various distributions using (\ref{eq:NonParametricVer})
 at the local node and (\ref{eq:FuseCUSUM}) at the fusion node: $EZ_k=-0.3$, $var(Z_k)=1$, $\rho=0.005$ and $b=1$.}}
\label{Tab:Pfa}
\end{table}
\subsection*{False Alarm outside a Batch}
\label{sec:Pfa_OutsideBatch}

In the absence of any transmission from the sensors, $Y_k \sim N(0, \sigma_{MAC}^2)$ 
if $Z_{MAC} \sim N(0, \sigma_{MAC}^2)$, where $N(0, \sigma_{MAC}^2)$ denotes Gaussian 
distribution with mean 0 and variance $\sigma_{MAC}^2$.
Hence, $F_k$ has negative drift. Thus the time to first reach $\beta$, i.e., time till FA, 
is approximately exponentially distributed with parameter $\lambda_0$ which can be 
obtained from Section \ref{sec:FPT}. 


\subsection{Comparative overall performance}
The effect of tail of $Z_k$ on FPT, overshoot and batch size was shown in the previous sections. 
This causes much larger $P_{FA}$ for heavy tailed $Z_k$ compared to the 
light tailed distributions for same mean and variance. This gets reflected into 
large $E_{DD}$ for heavy tailed distributions for a given $P_{FA}$.
Table \ref{Tab:EddComp} confirms these conclusions as the $E_{DD}$
for a light tailed system is much smaller as compared to the one from a heavy tailed system.
The individual systems 
are optimized to make sure that each performs at its best. 

Table \ref{Tab:EddCompPARANONPARA} shows the comparative performance of paramentric and nonparametric 
DualCUSUM's for given $f_0$ and $f_1$. The difference in performance is most pronounced when
the tail of $f_0$ is heaviest, i.e., for $K=7$, while the performance is same for Gaussian distributions on which
log likelihood function has no effect. 

Note that in Table \ref{Tab:EddCompPARANONPARA}, the variance of $Z_k$ is different for parametric and nonparametric 
CUSUMs. The overall effect is thus a combination of the effect of tails and that of the variances. However, as 
can be seen from the table, the effect of tail dominates and the general conclusion that light tailed 
systems are better, still holds.  

$E_{DD}$ in Tables \ref{Tab:EddComp} and \ref{Tab:EddCompPARANONPARA} are computed via simulations. 
However in the next section 
we theoretically evaluate $E_{DD}$ and then compare with the simulated values. 
\begin{table} [htbp]
\centering 
\centering
{\scriptsize
\begin{tabular}{|r|r|r|r|r|r|r|r|r|r|r|}
\hline 
$\rho$&$P_{FA}$&$E_{DD}^*$&$E_{DD}^*$\\
&&Gauss&Pareto\\
&&&K=2.1\\
&&&\\	
\hline
5e-4&e-2&29&36\\
5e-4&e-3&33&49\\
1e-4&e-3&41&95\\
\hline
\end{tabular} }
\makeatletter\def\@captype{table}\makeatother
\caption{{\scriptsize Comparison of $E_{DD}^*$ of Gaussian ($I^*=3$) and Pareto ($I^*=4$) 
with $L=5$, ${\cal E}_0 = 5$, $EZ_k=-0.5$, $var(Z_k)=1$ and $Z_{MAC}=N(0,1)$. }}
\label{Tab:EddComp}  
\end{table}

\vspace{-0.1cm}

\subsection{Computation of {\large $\textbf{\emph{E}}_{\textbf{\emph{DD}}}$}}
\label{sec:Analysis_Edd}

The detection delay, $E_{DD}$, at the fusion node, after the change has occurred, can be written as, 
\hspace{-0.5cm}
\begin{eqnarray}
E_{DD} = E\left[(\tau - T)^+\right] =  E[\tau-T | \tau \geq T ] (1-P_{FA}).
\end{eqnarray}

%
%

The $E_{DD}$ is atleast equal to the time it takes for some of the sensors to transmit and make the 
drift of $F_k$ positive. After that the additional delay is due to the time it takes for $F_k$ to cross 
$\beta$ with this positive drift, or due to the additional positive drift given by transmissions due to the
other remaining sensors. This concept is used to compute the $E_{DD}$ in the following.

When $\mu=EZ_k>0$, the time $\tau_\gamma$ for $W_k$ at a local node to cross 
threshold $\gamma$ satisfies, $E[\tau_\gamma]/\gamma \to 1/\mu$ as $\gamma\to \infty$.
Thus for large $\gamma$, $\tau_\gamma \sim \gamma/\mu$. Again for large $\gamma$, one can invoke Central 
Limit Theorem and approximate $\tau_\gamma$ by a Gaussian random variable:
 $\tau_\gamma \sim N(\frac{\gamma}{\mu}, \frac{\sigma^2\gamma}{\mu^3} )$,
where, $\sigma^2 = var(Z_k)$.

Let $\mu_l$ be the drift of fusion CUSUM $F_k$ when $l$ local nodes are transmitting. 

Let $t_i$ be the point at which the drift of $F_k$ changes from $\mu_{i-1}$ to $\mu_i$, i.e., 
when the drift of $F_k$ is already $\mu_{i-1}$ and one additional node (out of the remaining
$L-i+1$) starts transmitting. Call $t_i$ the $ith$ transition epoch.
Let $\tau_{x,1}, \tau_{x,2}, \ldots, \tau_{x,L}$ be iid with distribution of $\tau_x$. 
Then $t_1$, the time at which the first transmission happens, can be written as,
$t_1 = M(L, \gamma) = \min(\tau_{\gamma,1}, \tau_{\gamma,2}, \ldots, \tau_{\gamma,L}).$
Let $m(L, \gamma) = E[M(L, \gamma)]=E[t_1]$. Then 
$ E[t_2] \approx m(L, \gamma) + m(L-1, \gamma - m(L, \gamma)\mu).$
The term $m(L-1, \gamma - m(L, \gamma)\mu)$ represents 
the mean time for one of the remaining $L-1$ sensors to cross $\gamma$ given that their respective CUSUMs 
have run for $t_1$ slots, i.e., if $\gamma_{L-1} = \gamma - m(L, \gamma)\mu$, then 
$m(L-1, \gamma - m(L, \gamma)\mu) = E[\min (\tau_{\gamma_{L-1},1}, \tau_{\gamma_{L-1},2}, \ldots, 
\tau_{\gamma_{L-1},L-1})].$
In $t_1$ slots each of the $W_k$ has moved up by approximately
$E[t_1] \mu$ and hence the correction. This way we can compute the approximation to mean of each 
of the transition epochs: 
\[E[t_i] = \sum_{l=L}^i m(l, \gamma_l),\]
where, $\gamma_L=\gamma$ and 
$\gamma_l = \gamma_{l+1} - \mu. m(l+1, \gamma_{l+1})$ and $i\leq L$.
It may happen for large values of $\mu$ that, $m(k, \gamma_k) \leq 1;$ for some $k$. 
In that case we set $m(i,\gamma_i) = 0; k\geq i\geq 2$. 

\begin{table} [htbp]
\centering 
\centering
{\scriptsize
\begin{tabular}{|r|r|r|r|r|r|r|r|r|r|r|}
\hline 
$f_0 \to f_1$&$L/I$&$E_{DD}^*$&$E_{DD}^*$\\
&&nonparametric&parametric\\
\hline
Pareto $x_m=1$&5/4&54.8&4.4\\
$K=7$ to $K=3$&&&\\
\hline
Pareto $x_m=1$&5/4&69.1&24.9\\
$K=40$ to $K=30$&&&\\
\hline
Gaussian $\sigma=1$&5/3&10.1&10.1\\
$EZ_k=0$ to $EZ_k=0.6$&&&\\
\hline
\end{tabular} }
\makeatletter\def\@captype{table}\makeatother
\caption{{\scriptsize Comparative performance of parametric and nonparametric DualCUSUM for $P_{FA}=0.01$ 
with $\rho=0.05$, ${\cal E}_0 = 7.61$, and $Z_{MAC}=N(0,1)$.}}
\label{Tab:EddCompPARANONPARA}  
\end{table}
Now define, $\bar{F}_i = E[F_{t_i-1}]$, the mean value of $F_k$ just before the transition 
epoch $t_i$. Then a recursion can be written to approximately compute the $\bar{F}_i$:
\begin{eqnarray*}
\bar{F}_k = \bar{F}_{k-1} + 1_{\{\mu_{k-1} > 0\}} \left[\mu_{k-1} m(L-k+1, \gamma_{L-k+1})\right]; \ \ \ \ \bar{F}_1 = 0.
\end{eqnarray*}
Let, $j = \min \{i: \mu_i > 0 \mbox{ and } \frac{\beta-\bar{F}_i}{\mu_i} < E[t_{i+1}] \},$
then $E_{DD}$ can be approximated by, 
\begin{eqnarray}
\label{eq:EDDApprox}
E_{DD} \approx  E\left[ t_{j} \right] + \frac{\beta-\bar{F}_j}{\mu_j}, 
\end{eqnarray}
where, $m(k,x)$ can be computed as, $m(k,x)  = \sum_{i=0}^\infty P(\tau_x > i)^k.$


Since, the strong law of large number and the central limit theorem suffice
to build the approximations, the $E_{DD}$ is independent of the distribution of $Z_k$ 
but depends only on its mean and variance. 
The results are shown in Table \ref{Tab:Edd} for different distributions.  
The second column is our approximation (\ref{eq:EDDApprox}) and the rest are obtained via actual system 
simulations. It can be seen that, as $\gamma$ reaches 50, the $E_{DD}$ of all the distributions 
considered is nearly 147. 
\begin{table} [htbp]
\centering 
\centering
 {\scriptsize
 \begin{tabular}{|r|r|r|r|r|r|r|r|r|r|r|}
 \hline
 &&&&$E_{DD}$&$E_{DD}$\\
 $\gamma$&$E_{DD}$&$ E_{DD}$&$ E_{DD} $&$Log-$&$Pareto$\\
 &$Anal.$&$Gauss$&$Laplace$&$normal$&$K=3$\\
 \hline
 5&5.3&9.1&9.3&9.3&10.7\\
 8&11.4&16.6&16.8&16.9&18.7\\
 15&30.3&36.3&36.5&36.7&38.5\\
 50&146.7&146.8&147.1&147.6&150.5\\
 \hline
 \end{tabular} }
 \caption{{\scriptsize Comparison of $E_{DD}$ for various distributions: $L=10, I=1, \beta=\gamma$ $EZ_k=-0.3$, Var($Z_k$)=1 and $b=1$.}}
 \label{Tab:Edd}
\end{table}
\section{Conclusions}
\label{sec:conclusion} We have proposed an energy efficient 
distributed change detection scheme in \cite{icassp} which uses
the physical layer fusion technique and CUSUM at the sensors as well as 
at the fusion center. In this paper we extend the algorithm to also
include the nonparametric CUSUM. We have theoretically 
computed the probability of false alarm and mean delay in change detection.
The analytical results provide a good approximation which can then be used to choose the optimal parameters 
using the optimization technique in \cite{icassp}.

\end{document}